\documentclass{aa}

\usepackage{natbib}

\bibpunct{(}{)}{;}{a}{}{,} 

\usepackage{graphicx}

\usepackage{amssymb,amsmath}
\usepackage{delarray}
\usepackage{mathrsfs}
\usepackage{multirow}
\usepackage{color}
\usepackage{placeins}

\usepackage{balance}

\usepackage[pdftitle={Structure of the Balmer jump. The isolated hydrogen atom},%
            pdfauthor={Calvo F., Belluzzi L., Steiner O.},%
            pdfsubject={A\&A A613, A55 (2018). DOI: 10.1051/0004-6361/201731974},%
            pdfkeywords={atomic processes, opacity},%
            unicode=true,%
            bookmarks=true,%
            bookmarksnumbered=false,%
            bookmarksopen=false,%
            breaklinks=false,%
            pdfborder={0 0 .5},%
            backref=false,%
            colorlinks=true,%
            citecolor=blue,%
            linkcolor=blue,%
            urlcolor=blue]{hyperref}
\usepackage[all]{hypcap}

\usepackage[varg]{txfonts}

\usepackage{etoolbox}
\makeatletter

\patchcmd{\NAT@citex}
{\@citea\NAT@hyper@{%
                \NAT@nmfmt{\NAT@nm}%
                \hyper@natlinkbreak{\NAT@aysep\NAT@spacechar}{\@citeb\@extra@b@citeb}%
                \NAT@date}}
{\@citea\NAT@nmfmt{\NAT@nm}%
        \NAT@aysep\NAT@spacechar\NAT@hyper@{\NAT@date}}{}{}

\patchcmd{\NAT@citex}
{\@citea\NAT@hyper@{%
                \NAT@nmfmt{\NAT@nm}%
                \hyper@natlinkbreak{\NAT@spacechar\NAT@@open\if*#1*\else#1\NAT@spacechar\fi}%
                {\@citeb\@extra@b@citeb}%
                \NAT@date}}
{\@citea\NAT@nmfmt{\NAT@nm}%
        \NAT@spacechar\NAT@@open\if*#1*\else#1\NAT@spacechar\fi\NAT@hyper@{\NAT@date}}
{}{}

\makeatother

\makeatletter
\def\convertto#1#2{\strip@pt\dimexpr #2*65536/\number\dimexpr 1#1}
\makeatother

\newcommand{\vect}[1]{\boldsymbol{#1}}

\newcommand{\diff}{{\rm d}}

\newcommand{\threej}[6]{\left(\begin{array}{ccc}
                #1 & #2 & #3 \\
                #4 & #5 & #6
        \end{array}\right)}
\newcommand{\sixj}[6]{\left\{\begin{array}{ccc}
                #1 & #2 & #3 \\
                #4 & #5 & #6
        \end{array}\right\}}

\newcommand{\N}{\mathcal{N}}

\newcommand{\etaA}{\eta^\text{A}}

\newcommand{\Ju}{J_u}

\newcommand{\bbt}[1]{\left[#1\right]_{\rm bb}}
\newcommand{\bft}[1]{\left[#1\right]_{\rm bf}}
\newcommand{\reducedME}[3]{\left<#1\,\left\lVert\,#2\,\right\rVert\,#3\right>}
\newcommand{\braket}[3]{\left<#1\,\left|\,#2\,\right|\,#3\right>}

\newcommand{\Ll}{L_\ell}

\newcommand{\J}{\mathcal{J}}
\newcommand{\SetN}{\mathbb{Z}_{\geq 0}}

\begin{document}

\AANum{AA/2017/31974}

\title{\vspace{1.1cm}Structure of the Balmer jump}
\titlerunning{Structure of the Balmer jump. The isolated hydrogen atom}
\subtitle{The isolated hydrogen atom}

\author{F.~Calvo\inst{1,2}
        \and L.~Belluzzi\inst{1,3}
        \and O.~Steiner\inst{1,3}}

\institute{Istituto Ricerche Solari Locarno (IRSOL), 
        via Patocchi 57\,--\,Prato Pernice, 6605 Locarno-Monti, Switzerland\\
        \email{\href{mailto:flavio.calvo@irsol.ch}{\color{magenta}{flavio.calvo@irsol.ch}}}
        \and 
        Geneva Observatory, University of Geneva, 
        Ch. des Maillettes 51, 1290 Sauverny, Switzerland
        \and Kiepenheuer-Institut f\"ur Sonnenphysik, 
        Sch\"oneckstrasse 6, 79104 Freiburg, Germany
}

\date{Received 19 September 2017 / Accepted 3 December 2017}

\abstract
{The spectrum of the hydrogen atom was explained by Bohr more than one century ago. We revisit here some of the aspects of the underlying quantum structure, with a modern formalism, focusing on the limit of the Balmer series.
}
{We investigate the behaviour of the absorption coefficient of the isolated hydrogen atom in the neighbourhood of the Balmer limit.
}
{We analytically computed the total cross-section arising from bound-bound and bound-free transitions in the isolated hydrogen atom at the Balmer limit, and established a simplified semi-analytical model for the surroundings of that limit. We worked within the framework of the formalism of Landi Degl'Innocenti \& Landolfi (2004, Astrophys. Space Sci. Lib., 307), which permits an almost straight-forward generalization of our results to other atoms and molecules, and which is perfectly suitable for including polarization phenomena in the problem.
}
{We analytically show that there is no discontinuity at the Balmer limit, even though the concept of a ``Balmer jump'' is still meaningful. Furthermore, we give a possible definition of the location of the Balmer jump, and we check that this location is dependent on the broadening mechanisms. At the Balmer limit, we compute the cross-section in a fully analytical way.
}
{The Balmer jump is produced by a rapid drop of the total Balmer cross-section, yet this variation is smooth and continuous when both bound-bound and bound-free processes are taken into account, and its shape and location is dependent on the broadening mechanisms.
}


\maketitle

%
%
\section{Introduction}
\label{sec:intro}

This paper contains the first step of a work whose final goal is the 
numerical modelling of the Balmer jump in both the intensity and the 
linearly polarized spectrum of the solar radiation.
Here, we investigate the behaviour of the absorption coefficient
around the limit of the Balmer series, under the assumption of an isolated
hydrogen atom (i.e., an atom that does not interact with any other particle) 
and in the absence of magnetic fields.

A numerical calculation of the absorption coefficient of the hydrogen atom near the Balmer limit, 
including both bound-bound and bound-free processes, was 
carried out by \cite{2005A&A...429..713S}. There, the oscillator strengths 
(which are directly related to the Einstein $B$ coefficients) are computed from 
the Gaunt factors, which have historically been tabulated, and for which 
approximate formulas have been derived for quick computations. Thanks to the 
increased computational power available today, we have opted 
for a different approach in this work: we start from the analytical expression of the 
wavefunctions of electrons and compute oscillator strengths by evaluating 
 the radial integrals numerically. With this approach, we are no longer bound to 
existing tables or approximate formulas.

It must be observed that the radial integrals for the hydrogen atom have also 
been computed analytically, and the result expressed in a closed form 
\citep{1929AnP...394.1031G}. A generalized expression for hydrogenic atoms was 
later proposed by \citet{1935MNRAS..96...77M}, although without any proof. 
Almost thirty years later, \citet{1964RvMP...36..613M} provided the proof, in 
a simpler and more elegant way than in \citet{1929AnP...394.1031G}. A closed 
formula for the evaluation of the oscillator strengths has therefore long been 
available. It requires the evaluation of hypergeometric functions, however, which 
can be troublesome near the series limits. As we show
below, difficulties near the series limits are also met with the numerical 
integration methods that are developed in this work, but these methods have 
the advantage of being directly applicable to more complex electron wavefunctions.

The most exhaustive study of the hydrogen atom we found was performed in 
\cite{1957qmot.book.....B}, but we prefer to refer, whenever possible, to 
\cite{2004ASSL..307.....L} and more recent textbooks in order to keep a standard 
and more recent notation and formalism.

The structure of this contribution is as follows. In Sect.~\ref{sec:absorption_coeff} 
we provide the explicit expression of the bound-bound and bound-free absorption 
coefficients in terms of the density of quantum states and the Einstein $B$ 
coefficients. Sect.~\ref{sec:density_of_states} is devoted to the computation 
of the density of quantum states, whereas Sect.~\ref{sec:Einstein_coeffs} is 
dedicated to the formal computation of the Einstein coefficients from the 
analytic expression of the wavefunctions of the free and bound electrons. In 
Sect.~\ref{sec:absorption_coeff_limit} we match the absorption bound-bound 
and bound-free coefficients below and above the Balmer limit, respectively, and we 
analytically show that there is no discontinuity. In addition, we analytically 
compute the cross-section at that precise point. Sect.~\ref{sec:numerical_modelling} 
is a preliminary numerical calculation of the total cross-section (bound-bound and 
bound-free processes from the $n=2$ level) around the Balmer limit, taking only 
natural and thermal broadening into account.

\section{Absorption coefficient}
\label{sec:absorption_coeff}
We consider a multi-level atom, and we work within the framework of the 
formalism of \cite{2004ASSL..307.....L}. 
Each energy level is specified by the quantum numbers $(\alpha J)$, with $J$ 
the total angular momentum, and $\alpha$ a set of inner quantum numbers.
In the absence of magnetic fields and neglecting stimulated emission, the 
absorption coefficient due to 
bound-bound transitions is given by Eq.~(7.16a) of \cite{2004ASSL..307.....L}:
\begin{align}
        \bbt{\eta_i(\nu,\vect{\Omega})} = \ &
\frac{h\nu}{4\pi} \mathcal{N} \sum_{\alpha_{\ell} J_{\ell}} 
        \sum_{\alpha_u J_u} (2 J_\ell + 1) \, 
        B(\alpha_\ell J_\ell \rightarrow \alpha_u J_u) \nonumber \\
        & \times \sum_{KQ} \sqrt{3} \, (-1)^{1+ J_\ell + J_u + K}
        \sixj{1}{1}{K}{J_\ell}{J_\ell}{\Ju} \nonumber \\
        & \times \mathcal{T}^K_Q(i,\vect{\Omega}) \, \rho^K_Q(\alpha_\ell J_\ell)\, 
        \phi_{\rm bb}(\nu_{u \ell} - \nu),
        \label{Eq:abs_bb_pll_prof}
\end{align}
where $\nu$ is the frequency of the radiation, $\vect{\Omega}$ is the 
propagation direction, and where the index $i$ can take values 0, 1, 2, and 3,
standing for Stokes $I$, $Q$, $U$, and $V$, respectively. 
The quantity $h$ is the Planck constant, $\mathcal{N}$ is the number
density of atoms, $B(\alpha_\ell J_\ell \rightarrow \alpha_u J_u)$ is the
Einstein coefficient for absorption from a lower level $(\alpha_\ell J_\ell)$ 
to an upper level $(\alpha_u J_u)$, $\mathcal{T}^K_Q(i,\vect{\Omega})$ is the
polarization tensor \citep[see Table 5.6 in][]{2004ASSL..307.....L}, and 
$\rho^K_Q(\alpha_\ell J_\ell)$ are the 
multipolar components of the density matrix (or spherical statistical tensors)
of the lower level.
In the atomic reference frame, the absorption profile 
$\phi_{\rm bb}(\nu_{u \ell}-\nu)$ is a Lorentzian, with $\nu_{u \ell}$ the 
transition frequency defined by
\begin{equation}
        \nu_{u \ell} = \frac{E_u - E_{\ell}}{h},
\end{equation}
with $E_u$ and $E_\ell$ the energies of levels $(\alpha_u J_u)$ and
$(\alpha_\ell J_\ell)$, respectively.
The threshold frequency for photoionization from a given lower level is
\begin{equation}
        \nu_{\rm th}(\alpha_\ell J_\ell) = \frac{E_{\alpha_+ J_+} - 
        E_\ell}{h},
        \label{Eq:freq_thr}
\end{equation}
with $E_{\alpha_+ J_+}$ the energy of the level $(\alpha_+ J_+)$ in which the 
ion is left after photoionization.
Indicating with $\epsilon_u$ the (negative) energy of the upper level 
($\alpha_u J_u$) in a energy scale in which $E_{\alpha_+ J_+}=0$, the
transition frequency $\nu_{u \ell}$ can be written as
\begin{equation}
        \nu_{u \ell} = \nu_{\rm th}(\alpha_\ell J_\ell) + \frac{\epsilon_u}{h} \, .
\end{equation}

If the lower level is not polarized, then (see Eqs. (10.6) and (10.7) in 
\citealt{2004ASSL..307.....L} for more details)
\begin{equation}
        \rho^K_Q(\alpha_\ell J_\ell) = \delta_{K0} \, \delta_{Q0} 
        \frac{1}{\sqrt{2J_\ell + 1}} 
        \frac{\mathcal{N}_{\alpha_\ell J_\ell}}{\mathcal{N}},
        \label{Eq:rho_ull}
\end{equation}
with $\mathcal{N}_{\alpha_\ell J_\ell}$ the number density of atoms in the
lower level.
Substituting Eq.~(\ref{Eq:rho_ull}) into Eq.~(\ref{Eq:abs_bb_pll_prof}), 
observing that $\mathcal{T}^0_0(i,\vec{\Omega})=\delta_{i0}$, and using the
analytical expression of the 6-$j$ symbols when one of the arguments
is zero \citep[see Eq.~(2.36a) in][]{2004ASSL..307.....L}, we obtain the familiar expression
\begin{align}
        \bbt{\eta_i(\nu)} = & \, \delta_{i0} \,
        \frac{h\nu}{4\pi} \sum_{\alpha_{\ell} J_{\ell}} 
        \mathcal{N}_{\alpha_\ell J_\ell} \nonumber \\
        & \times 
        \sum_{\alpha_u J_u} B(\alpha_\ell J_\ell \rightarrow \alpha_u J_u) 
        \, \phi_{\rm bb} \left( \nu_{\rm th}(\alpha_\ell J_\ell) +
        \frac{\epsilon_u}{h} - \nu \right) \, .
        \label{Eq:abs_bb_prof}
\end{align}
As expected, in the absence of magnetic fields, and assuming that the lower 
level is unpolarized, only the absorption coefficient for the intensity 
$(i=0)$ is non-zero, and it does not depend on the propagation direction of the 
radiation.

The absorption coefficient for bound-free transitions (in the absence of 
atomic polarization in the lower level) can be easily obtained in a similar way
as Eq.~(\ref{Eq:abs_bb_prof}):
\begin{align}
        \bft{\eta_i(\nu)} = & \, \delta_{i0} \,
        \frac{h\nu}{4\pi} \sum_{\alpha_\ell J_\ell} 
        \mathcal{N}_{\alpha_\ell J_\ell} \nonumber \\
        & \times \int_0^\infty \diff \epsilon \mathcal{N}_{\rm f}(\epsilon) \, 
        \sum_{l j J'} 
        B(\alpha_\ell J_\ell \rightarrow \alpha_+ J_+, \epsilon l j, J') 
        \nonumber \\
        & \times \phi_{\rm bf} \left( \nu_{\rm th}(\alpha_\ell J_\ell) +
        \frac{\epsilon}{h} - \nu \right),
        \label{Eq:abs_bf_prof}
\end{align}
where $(\alpha_\ell J_\ell)$ is the bound level of the atom from which
photoionization takes place, $(\alpha_+ J_+)$ is the level at which the ion 
is left, $\epsilon$ is the (positive) energy of the released electron, $l$ 
and $j$ are its orbital and total angular momentum, respectively, and $J'$ 
is the total angular momentum of the final state 
($\vec{J'} = \vec{J_+} + \vec{j}$). 
The quantity $\mathcal{N}_{\rm f}(\epsilon)$ is the number density of quantum 
states of the free electron with energy $\epsilon$.
The profile $\phi_{\rm bf}$ can be defined similarly as the line profile 
$\phi_{\rm bb}$, but its exact shape is irrelevant in the following derivations.

We now particularize our formalism to the case of the hydrogen atom, which 
we describe neglecting the spin and relativistic corrections.
In this case, the atomic states are specified by the quantum numbers 
$(n l)$, with $n$ the principal quantum number ($n \ge 1$), and $l$ the 
orbital angular momentum (or azimuthal quantum number, $0\leq l \leq n-1$). 
The energies of the levels depend on the principal quantum number $n$ alone.
The degeneracy of the levels is $g(n)=n^2$.
Indicating with $(n l)$ the initial bound state, with $(n_u l_u)$ a given 
upper bound state, and with $(\epsilon l_f)$ an arbitrary free state with 
positive energy $\epsilon$, we have
\begin{align}
        \bbt{\eta(\nu)} = & \, \frac{h \nu}{4 \pi} \, 
        \sum_{n l} \mathcal{N}_{n l} \nonumber \\
        & \times \sum_{n_u l_u} B(n l \rightarrow n_u l_u) \, 
        \phi_{\rm bb} \left( \nu_{\rm th}(n) + \frac{\epsilon_u}{h} - 
        \nu \right),
        \label{eqn:absorption-coef-bb}
\end{align}
and
\begin{align}
        \bft{\eta(\nu)} = & \, \frac{h \nu}{4 \pi} \, 
        \sum_{n l} \mathcal{N}_{n l} \nonumber \\ 
        & \times \sum_{l_f} \int \diff \epsilon \, 
        \mathcal{N}_{\rm f}(\epsilon) \, B(n l \rightarrow \epsilon l_f) \, 
        \phi_{\rm bf} \left( \nu_{\rm th}(n) + \frac{\epsilon}{h} - 
        \nu \right),
        \label{eqn:absorption-coef-bf}
\end{align}
where we have dropped the index $i$, since only the absorption coefficient 
for the intensity is non-zero.

Near the photoionization limit, the bound states asymptotically approach the limit 
forming a quasi-continuum, so that the sum over the upper levels
$(n_u)$ in Eq.~(\ref{eqn:absorption-coef-bb}) can be formally substituted with an 
integral
\begin{align}
        \sum_{n_u} & \, B(n l \rightarrow n_u l_u) \, 
        \phi_{\rm bb} \left( \nu_{\rm th} + \frac{\epsilon_u}{h} - 
        \nu \right) \nonumber \\
        & \qquad 
        \rightarrow \int \diff \epsilon \, \mathcal{N}_{\rm b}(\epsilon) \, 
        \bar{B}(n l \rightarrow \epsilon l_u) \, 
        \phi_{\rm bb} \left( \nu_{\rm th} + \frac{\epsilon}{h} - 
        \nu \right),
        \label{eqn:integral_approx}
\end{align}
where $\mathcal{N}_b(\epsilon)$ is the number density of bound states with 
energy $\epsilon$, and $\bar{B}(n l \rightarrow \epsilon l_u)$ is the 
Einstein coefficient, defined over continuous values of the energy, for the 
bound-bound transition between the initial level $(n l)$ and the final level
with orbital angular momentum $l_u$ and energy $\epsilon$. 
This Einstein coefficient could be formally obtained by interpolating the Einstein
coefficient for the discrete spectrum of upper bound states, but a 
meaningful continuation of it will later appear in a natural way.

We now focus on the spectral region close to the photoionization
threshold from a given lower level $n$.
Close to the photoionization limit, in the frequency interval where the
absorption profiles $\phi_{\rm bb}$ and $\phi_{\rm bf}$ significantly contribute 
to the integral of Eq.~\eqref{eqn:integral_approx}, the integrand is practically constant. Anticipating the numerical calculations of Sect.~\ref{sec:numerical_modelling}, we note here that the variation of $\mathcal{N}_{\rm b}(\epsilon) \, \bar{B}(n l \rightarrow \epsilon l_u)$ across the wavelength range $3600$--$3700$~{\AA} is linear and as small as $10.5$\,\%. This variation is negligible as compared to the width of the profile, also in a realistic plasma, where due to electron-collisional broadening and other broadening mechanisms (Doppler broadening), the profile is expected to be significantly broader than that of the isolated hydrogen atom. The quantity $\mathcal{N}_{\rm b}(\epsilon) \, \bar{B}(n l \rightarrow \epsilon l_u)$ can then be evaluated at the wavelength corresponding to the maximum of the absorption profile $\phi_{\rm bb}$ and factorized out of the integral. We then obtain the following expressions for $[ \eta(\nu) ]_{\rm bb}$ and $[ \eta(\nu) ]_{\rm bf}$:
\begin{equation}
        \bbt{\eta(\nu)} = \frac{h^2 \nu}{4 \pi} \, 
        \sum_{l \, l_u} \mathcal{N}_{n l} \, \mathcal{N}_{\rm b}(\epsilon_0) \, 
        \bar{B}(n l \rightarrow \epsilon_0 l_u),
        \label{Eq:abs_bb_H}
\end{equation}
with $\epsilon_0 = h (\nu - \nu_{\rm th}) < 0$, and
\begin{equation}
        \bft{\eta(\nu)} = \frac{h^2 \nu}{4 \pi} \, 
        \sum_{l \, l_f} \mathcal{N}_{n l} \, \mathcal{N}_{\rm f}(\epsilon_0) \, 
        B(n l \rightarrow \epsilon_0 l_f),
        \label{Eq:abs_bf_H}
\end{equation}
with $\epsilon_0 = h (\nu - \nu_{\rm th}) > 0$.

Comparing Eqs.~(\ref{Eq:abs_bb_H}) and (\ref{Eq:abs_bf_H}), we see that 
close to the photoionization limit, the absorption coefficients for bound-bound 
and bound-free transitions have the same formal expression.
The next step is to derive and compare the explicit forms of the density of bound and 
free states ($\mathcal{N}_{\rm b}(\epsilon)$ and $\mathcal{N}_{\rm f}(\epsilon)$), 
and of the Einstein coefficients for bound-bound and bound-free transitions.

\section{Density of quantum states}
\label{sec:density_of_states}

On the one hand, the exact form of the density of bound states 
$\mathcal{N}_{\rm b}$ depends on the way the discrete spectrum of upper states is treated in a continuous manner. On the other hand, as we show below, the density of free states $\mathcal{N}_{\rm f}$ is related to the normalization of the wavefunction of the free electron. To calculate these quantities, as well 
as the Einstein coefficients, we work in Rydberg units. Then, the energy 
of the $n$th bound level of the hydrogen atom is given by
\begin{equation}
  \varepsilon_n = -\frac{1}{n^2},
\end{equation}
leading to
\begin{equation}
  \diff\varepsilon = \frac{2}{n^3}\,\diff n,
\end{equation}
and therefore
\begin{equation}
        \mathcal{N}_{\text{b}}(\varepsilon) \equiv \frac{\diff n}{\diff\varepsilon} = \frac{n^3}{2}.
        \label{eqn:n_bb}
\end{equation}

The determination of the density of quantum states in the free ``levels'' is more
involved, in the sense that the present formalism was developed for transitions
to discrete levels, and we are treating a continuous spectrum as a collection
of discrete levels in order to stick to the same mathematical tools.
Schematically, if we had split the energy range into smaller intervals and picked an 
approximative wavefunction of the free electron in each interval in order to compute 
the corresponding Einstein coefficient, the density of quantum states would depend on 
our splitting. In the limit where those intervals are infinitely small, the density of 
quantum states will depend on the way the wavefunction of the free electron is 
normalized. We later choose a normalization for which the density of quantum 
states is simply
\begin{equation}
        \mathcal{N}_{\text{f}}(\epsilon) = 1.
  \label{eqn:n_bf}
\end{equation}

\section{Einstein coefficients}
\label{sec:Einstein_coeffs}

The computation of the Einstein $B$ coefficients is split into two steps. First we 
express these coefficients as an integral over the wavefunctions of the bound and free 
electrons. Then we rewrite the wavefunctions of the bound and free electrons in a way 
that facilitates an analytical integration.

\subsection{Reduced matrix elements}

The Einstein coefficient for absorption from a lower state specified 
by the set of quantum numbers ($\beta_\ell$) to an upper state specified 
by the set $(\beta_u)$ is given by \citep[e.g. combine Eqs. (7.6) and (7.7) 
in][]{2004ASSL..307.....L}
\begin{equation}
        B(\beta_\ell \rightarrow \beta_u) = \frac{32 \pi^4}{3 h^2 c} \, 
        \left| \reducedME{\beta_\ell}{\vec{d}}{\beta_u} \right|^2,
        \label{eqn:bf-einstein-coef}
\end{equation}
where $\vec{d}=-e_0\vec{r}$ is the dipole operator, and 
$\reducedME{\cdot}{\vec{d}}{\cdot}$ is the corresponding reduced matrix element.
In our case, the bottom line of the problem restricts to the evaluation of the 
reduced matrix elements $\reducedME{n l}{\vec{d}}{n_u l_u}$ (for bound-bound
transitions), and $\reducedME{n l}{\vec{d}}{\epsilon l_f}$ (for bound-free 
transitions).
As shown in detail in Appendix \ref{sec:reduced_matrix_elements}, the reduced matrix 
element for bound-bound transitions is given by
\begin{align}
        \reducedME{n l}{\vec{d}}{n_u l_u} = & \, e_0 (-1)^{l_u+1}
        \sqrt{(2 l_u +1)} \nonumber \\
        & \times \threej{l}{l_u}{1}{0}{0}{0} \, 
        \mathcal{I}(n l, n_u l_u), 
        \label{eqn:reduced_matrix_bound}
\end{align}
where $\mathcal{I}$ is the radial integral defined by
\begin{equation}
        \mathcal{I}(n l, n_u l_u) = \int_0^\infty P_{n l}(r) \, r \, 
        P_{n_u l_u}(r) \, \diff r,
        \label{eqn:radial_integral_bound}
\end{equation}
with $P_{n l}$ the reduced radial wavefunction of the bound electron.
Similarly, for the bound-free transitions, we have
\begin{align}
        \reducedME{n l}{\vec{d}}{\epsilon l_f} = & \, e_0 (-1)^{l_f+1}
        \sqrt{(2 l_f +1)} \nonumber \\
        & \times \threej{l}{l_f}{1}{0}{0}{0} \, 
        \mathcal{I}(n l, \epsilon l_f), 
        \label{eqn:reduced_matrix_free}
\end{align}
with
\begin{equation}
        \mathcal{I}(n l, \epsilon l_f) = \int_0^\infty P_{n l}(r) \, r \, 
        \chi_{\epsilon l_f}(r) \, \diff r,
        \label{eqn:radial_integral_free}
\end{equation}
where $\chi_{\epsilon l_f}(r)$ is the reduced radial wavefunction of the 
free electron.

\subsection{Radial wavefunctions}
\label{sec:radial_wavefunctions}

The normalized radial wavefunctions are found by solving the radial Schrödinger 
equation for the potential $V(\xi)=-\frac{2}{\xi}$,
\begin{equation}
  \left(\frac{\diff^2}{\diff \xi^2}-\frac{l(l+1)}{\xi^2}-V(\xi)+\varepsilon\right)\, y=0,
  \label{eqn:radial-schrodinger-equation}
\end{equation}
%
%
which has been nondimensionalized (or rewritten in Rydberg units) using the 
substitutions $r=a_0\,\xi$ and $\epsilon=e_0^2\varepsilon/(2a_0)$ with 
$a_0=h^2/(4\pi^2m_ee_0^2)$ the Bohr radius. Taking into account the normalization
\begin{equation}
\int_{\xi=0}^{\infty} |y(\xi)|^2\diff\xi = \int_{r=0}^{\infty} |\tilde{y}(r)|^2\diff r = 1,
\end{equation}
we obtain the transformation law of the wavefunction $y$ under change of variables,
\begin{equation}
y(\xi) = \sqrt{a_0}\tilde{y}(r).
\end{equation}
Hence the radial integrals \eqref{eqn:radial_integral_bound} and 
\eqref{eqn:radial_integral_free} are related to the radial integral in Rydberg units
\begin{equation}
\begin{split}
\mathcal{J}(n l, n_u l_u) &\equiv \int_0^\infty P_{n l}(\xi) \, \xi \, 
P_{n_u l_u}(\xi) \, \diff\xi, \\
\mathcal{J}(n l, \varepsilon l_f) &\equiv \int_0^\infty P_{n l}(\xi) \, \xi \, 
\chi_{\varepsilon l_f}(\xi) \, \diff\xi,
\end{split}
\label{eqn:radial_integrals_nondimensional}
\end{equation}
by the transformations
\begin{equation}
\begin{split}
\mathcal{I}(n l, n_u l_u) &= a_0\,\mathcal{J}(n l, n_u l_u), \\
\mathcal{I}(n l, \epsilon l_f) &= a_0\,\mathcal{J}(n l, \varepsilon l_f).
\end{split}
\end{equation}

For $\varepsilon=-1/n^2<0$, the solutions of the radial Schrödinger 
Eq.~\eqref{eqn:radial-schrodinger-equation} can be found in \citet{1960MNRAS.120..121B} 
or in \citet{1995iqm..book.....G}:
\begin{equation}
\begin{split}
P_{nl}(\xi) &= \frac{1}{n}\frac{(-1)^{n+l+1}}{\sqrt{\,\varGamma(n+l+1)\,\varGamma(n-l)}}W_{n,l+\frac{1}{2}}\left(\frac{2\xi}{n}\right) \\
&= \frac{1}{n}\sqrt{\frac{\varGamma(n-l)}{\varGamma(n+l+1)}} \left(\frac{2\xi}{n}\right)^{l+1}\exp\left(-\frac{\xi}{n}\right)L_{n-l-1}^{(2l+1)}\left(\frac{2\xi}{n}\right),
\end{split}
\end{equation}
which are either expressed in terms of the Whittaker $W$ function or in terms of the 
generalized Laguerre polynomial (whose normalization differs in the literature),
\begin{equation}
L_n^{(\alpha)}(z) = \frac{\varGamma(n+\alpha+1)}{\varGamma(\alpha+1) \varGamma(n+1)}
\,_1F_1(-n, \alpha+1, z),
\end{equation}
\citep[see e.g. ][]{abramowitz1964handbook}. In turn, the confluent hypergeometric 
function $_1F_1$ can be expressed through the Whittaker $M$ function:
\begin{equation}
M_{\kappa,\mu}(z) = \exp\left(-\frac{z}{2}\right)z^{\mu+\frac{1}{2}}\,_1F_1\left(\mu-\kappa+\frac{1}{2},1+2\mu;z\right),
\end{equation}
allowing us to rewrite $P_{nl}$ in terms of the Whittaker $M$ function:
\begin{equation}
\begin{split}
P_{nl}(\xi) &= \sqrt{\frac{\varGamma(n+l+1)}{n^{2l+1}\varGamma(n-l)}}\frac{\sqrt{n^{2l-1}}M_{n,l+\frac{1}{2}}\left(\frac{2\xi}{n}\right)}{\varGamma(2l+2)} \\
&\equiv g_{nl}\ \frac{\sqrt{n^{2l-1}}M_{n,l+\frac{1}{2}}\left(\frac{2\xi}{n}\right)}{\varGamma(2l+2)}.
\end{split}
\label{eqn:P_nl}
\end{equation}
We note that the dependence of the Einstein coefficient $B(n l \rightarrow n_u l_u)$ 
for bound-bound transitions on the $n_u$ quantum number of the upper state appears 
only through the Whittaker $M$ function, hence providing a natural analytical 
continuation over non-integer values of $n_u$ that we have previously introduced as 
$\bar{B}(n l \rightarrow n_u l_u)$.

For $\varepsilon>0$, the solutions of the radial Schrödinger 
Eq.~\eqref{eqn:radial-schrodinger-equation} are listed in \cite{1958MNRAS.118..504S}. A 
unique combination of these functions with the appropriate boundary conditions 
for the present problem, namely that the wavefunction reduces to zero in the 
origin and its asymptotic form at large $r$ behaves as the wavefunction of a 
free electron, is presented here:
\begin{align}
        \chi_{\varepsilon l}(\xi) &= \frac{1}{\kappa^{l+1}}\!
        \sqrt{\frac{ \varGamma(l+1+i\kappa)\varGamma(l+1-i\kappa)}
        {\varGamma(i\kappa)\varGamma(-i\kappa)(1-\exp(2\pi\kappa))}}
        \frac{(i\kappa)^{l+1}M_{i\kappa,l+\frac{1}{2}}
        \left(-\frac{2i\xi}{\kappa}\right)}{\sqrt{2}\varGamma(2l+2)} 
        \nonumber \\
        & \equiv h_{\kappa l}\ \frac{(i\kappa)^{l+1}M_{i\kappa,l+
        \frac{1}{2}}\left(-\frac{2i\xi}{\kappa}\right)}{\sqrt{2}\varGamma(2l+2)},
\label{eqn:chi_el}
\end{align}
where $\kappa\equiv1/\sqrt{\varepsilon}$.

\section{Total absorption coefficient at the Balmer limit}
\label{sec:absorption_coeff_limit}

\subsection{Continuity of the total absorption coefficient}

Using the properties of the Euler gamma function, it is straightforward to 
check that
\begin{equation}
\lim_{n_u \rightarrow \infty} g_{n_u l_u} = \lim_{\kappa \rightarrow \infty}
h_{\kappa l_f} = 1 \quad \forall l_u = l_f \in \SetN.
\end{equation}

In Sect.~\ref{sec:radial_wavefunctions} we have noted that the Einstein coefficient 
$B(n l \rightarrow n_u l_u)$ is also well defined for non-integer values of $n_u$, and 
in particular, that it is a continuous function of $n_u$. Hence
\begin{equation}
\lim_{\varepsilon\rightarrow 0^-} \mathcal{N}_{\text{b}}(\varepsilon) 
\bar{B}(n l \rightarrow n_u(\varepsilon) l_u) = \lim_{n_u\rightarrow\infty} 
\mathcal{N}_{\text{b}}(\varepsilon(n_u)) 
B(n l \rightarrow n_u l_u).
\end{equation}
Now recalling the relation \eqref{eqn:whitM-lim}, namely
\begin{equation}
\lim_{\kappa\rightarrow\infty} \kappa^{\mu+\frac{1}{2}}M_{\kappa,\mu}\left(\frac{z}{\kappa}\right) = \lim_{\kappa\rightarrow\infty} (i\kappa)^{\mu+\frac{1}{2}} M_{i\kappa,\mu}\left(-\frac{iz}{\kappa}\right),
\end{equation}
which is demonstrated in Appendix \ref{sec:whitM_asympt_prop}, and using Eq. 
\eqref{eqn:P_nl} for the wavefunction of the bound electron $P_{nl}$ and Eq. 
\eqref{eqn:chi_el} for the wavefunction of the free electron $\chi_{\varepsilon\ell}$, 
we find that
\begin{equation}
        \lim_{\varepsilon\rightarrow 0^-} \mathcal{N}_{\text{b}}(\varepsilon) 
        \bar{B}(n l \rightarrow n_u l_u) = \lim_{\varepsilon\rightarrow 0^+}
        \mathcal{N}_{\text{f}}(\varepsilon) B(n l \rightarrow \varepsilon l_f),
        \label{eqn:NB-limits}
\end{equation}
provided that the density of quantum states are taken to be the ones obtained in 
Eqs. \eqref{eqn:n_bb} and \eqref{eqn:n_bf}. The equality of Eq.~\eqref{eqn:NB-limits}
further implies the interchange of the limits with the radial integrals contained 
in the Einstein $B$ coefficients, whose justification is similar to the proof of Eq. 
\eqref{eqn:whitM-lim}. For brevity, we do not repeat all the computations here. Some 
additional details are provided in the final note of Appendix 
\ref{sec:whitM_asympt_prop}, however.

After summing the bound-bound contributions to Eq.~(\ref{Eq:abs_bb_H}) over $l$ 
and $l_u$ and the bound-free contributions to Eq.~(\ref{Eq:abs_bf_H}) over $l$ 
and $l_f$, we obtain our final result,
\begin{equation}
        \lim_{\nu \rightarrow \nu_{\rm th}^-} \left[ \eta(\nu) \right]_{\rm bb} = 
        \lim_{\nu \rightarrow \nu_{\rm th}^+} \left[ \eta(\nu) \right]_{\rm bf},
\end{equation}
with the threshold frequency for the Balmer series given by
\begin{equation}
\nu_{\rm th} = \frac{m_ee_0^4}{16\pi\hbar^3}.
\label{eqn:balmer-freq}
\end{equation}
Thus, we have proven that there is no discontinuity at the Balmer limit.

\subsection{Analytic expression for the total absorption coefficient}

The wavefunction of the free electron with zero kinetic energy can be calculated 
combining Eqs. \eqref{eqn:chi_el} and \eqref{eqn:whitM-lim}:
\begin{equation}
\chi_{\varepsilon=0,\ell}(r) = \frac{(2r)^{\ell+1}\,_0F_1(2\ell+2;-2r)}{\sqrt{2}\varGamma(2\ell+2)}.
\label{eqn:limi_wavefunction_free}
\end{equation}
The integrals of Eqs. \eqref{eqn:radial_integral_bound} and 
\eqref{eqn:radial_integral_free} can be computed analytically. If we take the 
limiting wavefunction for the free electron, Eq. \eqref{eqn:limi_wavefunction_free}, 
and we consider the bound electron to be in the level $n=2$, we find the 
nondimensional expressions of Eq. \eqref{eqn:radial_integrals_nondimensional}:
\begin{equation}
\begin{split}
\J(n=2,L_\ell=0;\varepsilon=0,\ell=1) &= \frac{2^8}{e^4}, \\
\J(n=2,L_\ell=1;\varepsilon=0,\ell=0) &= \frac{2^7}{\sqrt{3}e^4}, \\
\J(n=2,L_\ell=1;\varepsilon=0,\ell=2) &= \frac{2^9}{\sqrt{3}e^4}.
\end{split}
\end{equation}
These are the only non-vanishing integrals, according to the selection rule 
$\Delta\ell=\pm 1$. Now inserting the reduced matrix 
\eqref{eqn:reduced_matrix_free} into Eq.~\eqref{eqn:bf-einstein-coef}, we compute 
the Einstein coefficients (where we have substituted the corresponding 
nondimensional radial integrals $\J$):
\begin{equation}
B(n \Ll \rightarrow \varepsilon\ell) = \frac{2h^2}{3ce_0^2m_e^2}\,(2\ell+1)\, \threej{\Ll}{\ell}{1}{0}{0}{0}^2\ \J^2(n\Ll,\varepsilon\ell).
\end{equation}
For $n=2$ we can sum over all values of all contributions from the angular 
momentum of the bound and free electrons, in order to obtain the absorption 
coefficient at the Balmer limit. Next, we note that the 3-$j$ symbol for 
$\ell=L_\ell\pm1$ can be computed using the 
property 
\begin{equation}
\threej{a}{a+1}{1}{\alpha}{-\alpha}{0} = (-1)^{a-\alpha-1}\sqrt{\frac{(a-\alpha+1)(a+\alpha+1)}{(a+1)(2a+1)(2a+3)}},
\end{equation}
and the fact that squares of 3-$j$ symbols are invariant under permutation of 
any two columns:
\begin{equation}
(2\ell+1)\, \threej{\Ll}{\ell}{1}{0}{0}{0}^2 = \frac{\max(L_\ell,\ell)}{2L_\ell+1}.
\end{equation}
Making use of Eqs. \eqref{eqn:absorption-coef-bb} and  
\eqref{eqn:absorption-coef-bf}, summing over the final states, and converting 
the density of quantum states into c.g.s units finally leads us to
\begin{equation}
\etaA_0(\nu_{\rm th}) = \frac{h^6\nu_{\rm th}}{12\pi^3 ce_0^6m_e^3} \left(\N_{n=2;\,L_\ell=0}\frac{2^{16}}{e^8}+\N_{n=2;\,L_\ell=1}\left(\frac{2^{14}}{9e^8}+\frac{2^{19}}{9e^8}\right)\right).
\end{equation}
Using the Boltzmann equation, we obtain the relative populations of the $n=2$ level:
\begin{equation}
\frac{\N_{n=2;\,L_\ell=0}}{\N_{n=2}} = \frac{1}{4},\quad\frac{\N_{n=2;\,L_\ell=1}}{\N_{n=2}} = \frac{3}{4},
\label{eqn:relative-populations}
\end{equation}
providing the final result:
\begin{equation}
\begin{split}
\etaA_0(\nu_{\rm th}) &= \N_{n=2}\frac{h^6\nu_{\rm th}(n=2)}{12\pi^3 ce_0^6m_e^3}\frac{2^{12}\times 3\times 5}{e^8} \\
&= \N_{n=2}\frac{4\pi^2}{3}\left(\frac{\alpha a_0^2}{n^2}\right)_{n=2}\frac{2^{12}\times 3\times 5}{e^8} \\
&\approx \N_{n=2} \times 1.386 \times 10^{-17}\,[\text{cm}^2],
\end{split}
\end{equation}
in which we have introduced the frequency at the Balmer limit given in 
Eq.~\eqref{eqn:balmer-freq}, and where $a_0$ is the Bohr radius and 
$\alpha=e_0^2/(\hbar c)\approx 1/137$ is the fine structure constant.

\section{Numerical modelling of the total cross-section around the Balmer limit}
\label{sec:numerical_modelling}

At some point near the Balmer limit, the distance between consecutive spectral lines 
becomes smaller than the width of the lines (FWHM). The exact 
location of this wavelength $\lambda_t$ representing the transition between these two 
regimes depends on the broadening of the spectral lines. At wavelengths 
longer than $\lambda_t$ , we still observe distinct spectral lines, but at shorter 
wavelengths, we no longer resolve spectral lines, but observe a 
quasi-continuum.

\begin{figure}[ht!]
	\centering
	\includegraphics[width=.95\columnwidth]{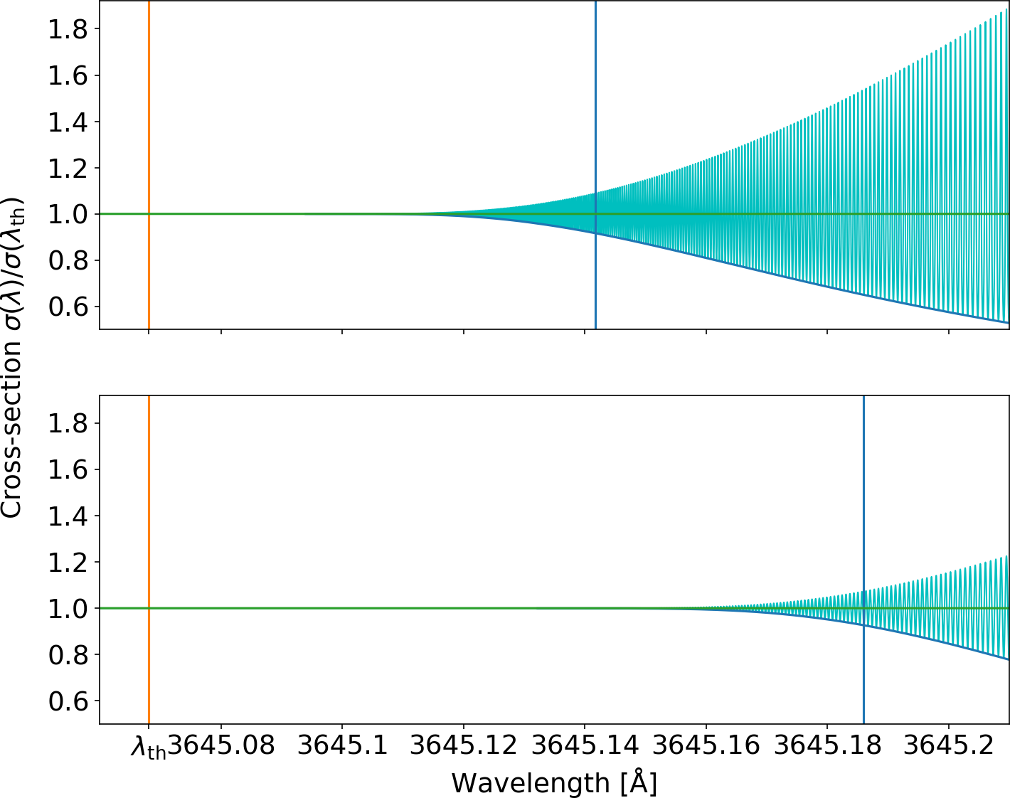}
	\caption{Total cross-section (cyan) due to bound-bound and bound-free transitions 
		in the isolated hydrogen atom around the Balmer limit $\lambda_\mathrm{th}$, 
		considering natural broadening (top), or including both natural 
		broadening and Gaussian broadening for a thermal velocity of 
		$15\,\textrm{m}\,\textrm{s}^{-1}$ (bottom). The cross-section is normalized 
		to its analytic value at the Balmer limit $\lambda_{\mathrm{th}}$. At 
		longer wavelengths, the actual continuum (blue curve), defined as the lower 
		envelope of the bound-bound cross-section, deviates from the nominal 
		quasi-continuum (almost-horizontal green curve that was computed by assuming 
		the approximation of Eq.~\eqref{eqn:integral_approx} to be valid at all 
		wavelengths $\lambda>\lambda_\mathrm{th}$). The orange vertical solid lines show 
		the location of the Balmer limit $\lambda_\mathrm{th}$. The blue vertical 
		lines show the location at which the FWHM of spectral lines and the 
		separation between consecutive spectral lines are equal ($\lambda_t$). In 
		the bottom plot, this line is displaced to longer wavelength due to the 
		additional broadening caused by the thermal velocity.}
	\label{fig:bj-cross-section}
\end{figure}

\begin{figure*}
	\centering
	\includegraphics[width=0.8\textwidth]{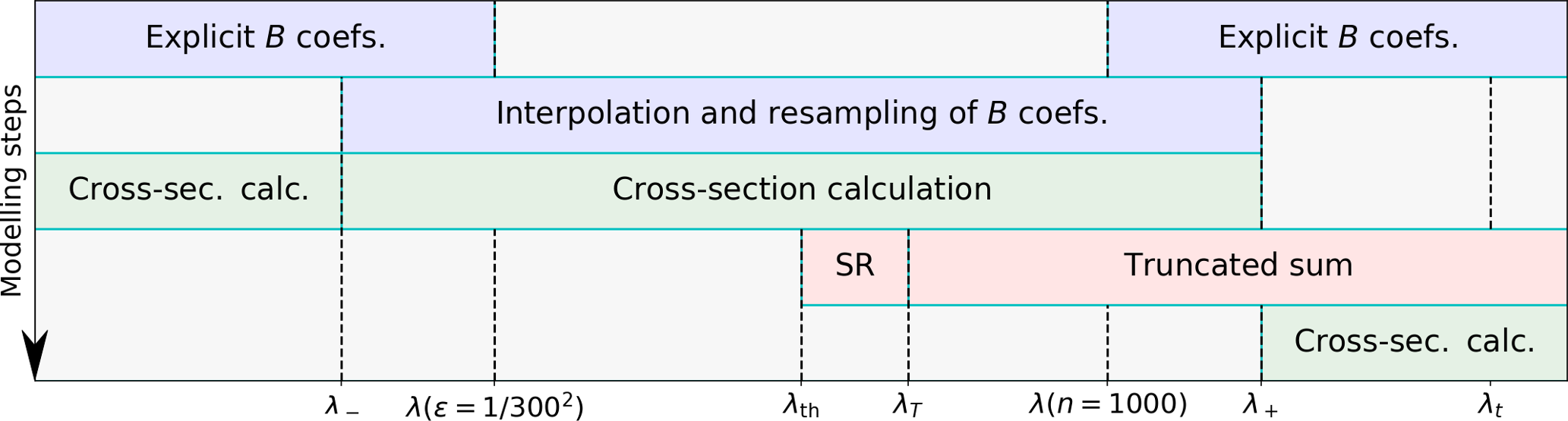}
	\caption{Steps involved in the modelling of the total cross-section around 
		the Balmer limit. The Einstein $B$ coefficients are first calculated, 
		either explicitly or by interpolation (blue). The cross-section is then 
		computed for $\lambda<\lambda_+$ (upper green region) using either 
		Eq.~\eqref{Eq:abs_bb_H} or Eq.~\eqref{Eq:abs_bf_H}. For $\lambda>\lambda_+$, 
		the distance between lines is comparable to the line's FWHM, so that 
		the approximation of Eq.~\eqref{eqn:integral_approx} is no longer valid, and 
		lines have to be computed individually, considering both a finite truncated 
		sum and its sum remainder (SR; red). The opacity of the lines is finally 
		summed at wavelengths $\lambda>\lambda_+$ (lower green region).}
	\label{fig:modelling}
\end{figure*}

In the following paragraphs we describe the procedure we used to numerically compute 
the total cross-section in the vicinity of the Balmer limit 
(Fig.~\ref{fig:bj-cross-section}). The total Balmer cross-section is defined by
\begin{equation}
\begin{split}
\sigma_\mathrm{Bal.}(\nu) = \frac{h^2 \nu}{4 \pi} \, \sum_{\ell} \left(\frac{\mathcal{N}_{n=2;\, l=0}}{\mathcal{N}_{n=2}} \, \right. & \bar{B}(n=2;l=0 \rightarrow \epsilon_0 \ell) \\
& \left. \!\!\!\!\!\! + \frac{\mathcal{N}_{n=2;\, l=1}}{\mathcal{N}_{n=2}} \, \bar{B}(n=2;l=1 \rightarrow \epsilon_0 \ell) \right),
\end{split}
\end{equation}
where $\ell$\ stands here for either $l_u$ or $l_f$, and the relative populations 
of the $n=2$ level are taken to be the same as in 
Eq.~\eqref{eqn:relative-populations}, with $l$ being a shorthand notation for 
the angular momentum of the lower lever $L_\ell$. The procedure to compute this 
cross-section is non-trivial because the wavefunctions of both
the free and bound 
electrons have singularities at the Balmer limit, leading to numerical 
difficulties around this region. Moreover, the number of spectral lines to be 
considered is in principle infinite.
The total cross-section at wavelengths larger than the wavelength of the Balmer 
limit ($\lambda_\mathrm{th}$) is given as an infinite sum of cross-sections due to 
each transition from the second level to any higher level. Approaching (from 
larger wavelengths) the specific point where the line width and the spacing 
between neighbouring lines are comparable ($\lambda=\lambda_t$), this sum cannot 
be truncated because the remaining terms are non-negligible and their sum 
decreases extremely slowly when the truncation threshold increases. However, it 
is possible to split the sum into two terms, one containing a sum over a finite 
number of spectral lines located at wavelengths longer than some threshold, 
$\lambda_T$, and one containing an infinite sum over spectral lines located at 
shorter wavelengths (between $\lambda_\mathrm{th}$ and $\lambda_T$). The infinite 
sum can then be approximated by an integral, with the same approach as used in 
Eq.~\eqref{eqn:integral_approx}.

The choice of $\lambda_T$ is quite empirical, but 
$\lambda_T\ll\lambda_t$ is always required and $\lambda_T$ has therefore to be 
adjusted according to the line broadening. In the limit 
$\lambda_T\rightarrow\lambda_\mathrm{th}$, where the chosen threshold approaches 
the Balmer limit, the error arising from the approximation in 
Eq.~\eqref{eqn:integral_approx} drops to zero, but the number of lines 
for which the Einstein coefficient has to be explicitly computed increases drastically. 
Moreover, close to the limit $\lambda_\mathrm{th}$, both below and above it, 
numerical evaluation at finite precision of the Einstein $B$ coefficients is 
troublesome \citep[see e.g.][]{2014MNRAS.441.2855M}. For simplicity, in the 
present work, we explicitly compute the Einstein $B$ coefficients inside a safe 
range (for $1000$ bound-bound transitions and $300$ bound-free transitions) and 
interpolate its values for transitions closer to the limit. More precisely, we 
interpolate the values of 
$\mathcal{N}_{\text{b}}(\varepsilon) B(n l \rightarrow n_u l_u)$ 
together with those of 
$\mathcal{N}_{\text{f}}(\varepsilon) B(n l \rightarrow \varepsilon l_f)$, 
since we know, according to Eq.~\eqref{eqn:NB-limits}, that they match at the 
Balmer limit. The evaluation of the Einstein $B$ coefficients is illustrated in 
blue in Fig.~\ref{fig:modelling}.

The wavelength range in which we perform the numerical calculation of the total 
cross-section is finally split into three sub-domains, represented in green 
in Fig.~\ref{fig:modelling}: $\lambda < \lambda_-$, for which 
Eq.~\eqref{Eq:abs_bf_H} is used; 
$\lambda_- \leq \lambda < \lambda_+$, for which either Eq.~\eqref{Eq:abs_bb_H} 
or Eq.~\eqref{Eq:abs_bf_H} are used, but the Einstein $B$ coefficients have 
interpolated and resampled values; and $\lambda \geq \lambda_+$, for which the 
sum in Eq.~\eqref{eqn:absorption-coef-bb} is truncated. The truncated part 
includes a total of $5000$ bound-bound transitions, most of which were computed 
with the interpolated $B$ coefficients, and the remainder of the sum is 
approximated using Eq.~\eqref{eqn:integral_approx}, in which all $B$ coefficients 
are interpolated (SR). This is schematically represented by the red domains in 
Fig.~\ref{fig:modelling}.

We note that for $\lambda<\lambda_\mathrm{th}$, the total cross-section 
varies very slowly, so that the choice of $\lambda_-$ is not really important 
as the interpolation between $\lambda_-$ and $\lambda_\mathrm{th}$ will provide 
accurate results. However, $\lambda_T$ should be chosen as close as possible to 
$\lambda_\mathrm{th}$ and should satisfy both $\lambda_T\ll\lambda_t$ and 
$\lambda_T\ll\lambda_+$ to obtain an accurate total cross-section, requiring an 
explicit computation of a large number of lines. The considered $1000$ 
bound-bound transitions with the Einstein $B$ coefficients computed 
explicitly is not sufficient, and additional transitions (till $n_u=5000$, 
corresponding to $\lambda_T$) are considered using the interpolated 
Einstein $B$ coefficients.

\begin{figure}
	\centering
	\includegraphics[width=0.95\columnwidth]{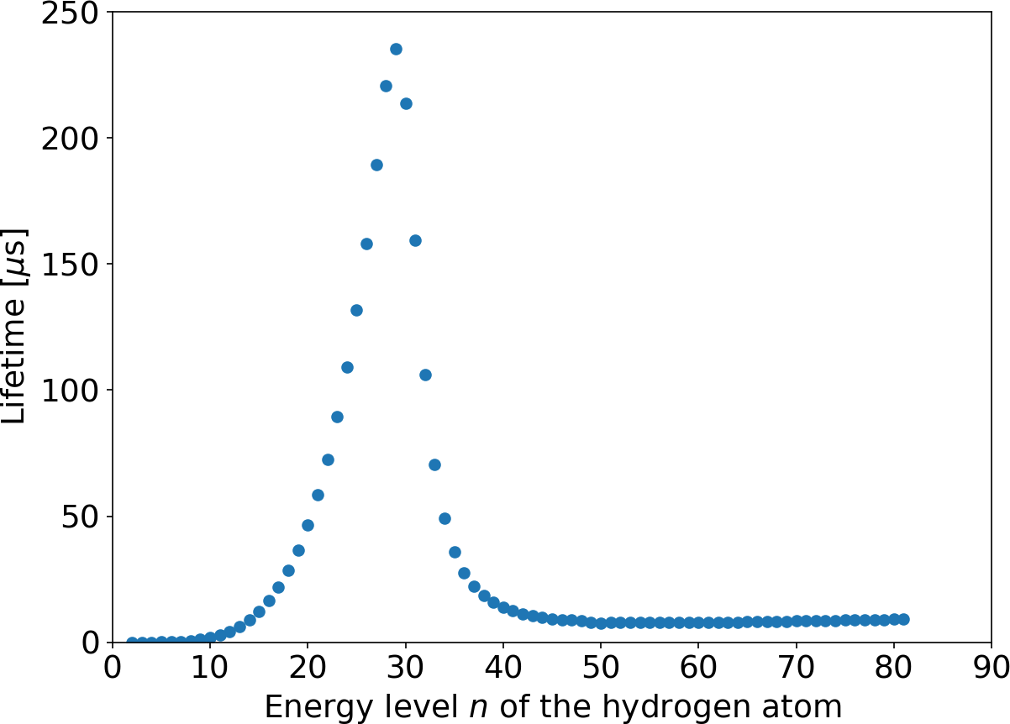}
	\caption{Lifetimes of the atomic levels $n$ of the hydrogen atom for $1 < n < 82$.}
	\label{fig:lifetimes}
\end{figure}

For line transitions to a higher level with quantum number $n_u$ above the $n_u=81$ 
threshold, the natural broadening of the lines is extrapolated. Below that 
threshold, the natural broadening is explicitly computed using Eq.~(6.59b) in 
\cite{2004ASSL..307.....L} (see Fig.~\ref{fig:lifetimes}):
\begin{equation}
\varGamma = \frac{\gamma_\textrm{l} + \gamma_\textrm{u}}{4\pi},
\end{equation}
where $\gamma_\textrm{l}$ and $\gamma_\textrm{u}$ are the inverse lifetimes of the 
lower and upper levels. The inverse lifetimes are given by the effective Einstein 
$A$ coefficients, which are computed using Eq.~(63.8) in \cite{1957qmot.book.....B} 
and the usual relations between the Einstein $A$ and $B$ coefficients. Above 
$n_u=81$, the extrapolation is carried out by fitting a (semi-empirical) power law 
$\gamma_u(n)\sim a n^{-b}$.

Figure~\ref{fig:bj-cross-section} was produced using the methods described above. It 
is not to be directly compared to a realistic plasma, but it exhibits some 
characteristic features that are expected to be found in observations
as well. First of all, 
no discontinuity is found anywhere, and in particular, nothing special occurs at 
the Balmer limit.
However, the lower envelope of the spectral lines, interpreted as the ``continuum'', 
quickly drops to lower cross-sections. On a spectrum exhibiting a wider 
range of wavelengths, it would almost appear as a ``jump''.

The location of the jump is not well defined. Longward of the Balmer limit, 
close inspection indeed reveals the existence of oscillations up to the limit. With a 
limited spectral resolution, however, there will be a wavelength shortward of which 
the oscillating cross-section (lines) become a quasi-continuum. In 
Fig.~\ref{fig:bj-cross-section} the vertical blue line on the right of each panel 
is the position at which the FWHM of spectral lines is equal to the line 
separation, and it could be taken as the definition of the Balmer's jump position.

By comparing the plot in the top panel of Fig.~\ref{fig:bj-cross-section} (no thermal 
broadening) with the plot in the bottom panel (thermal broadening corresponding to 
a thermal velocity of $15\,\textrm{m}\,\textrm{s}^{-1}$), we see that the 
(rather arbitrary) point in which the lines become a quasi-continuum is displaced 
towards longer wavelengths. The position of this point is very sensitive to the 
broadening mechanisms that are considered. In the solar plasma, with a temperature of 
about $6200\,\mathrm{K}$ (as expected at the depth at which the solar continuum is 
formed), and considering only natural and thermal broadening, we expect a displacement 
of the Balmer jump of approximately $7\,\AA$.

We note that an alternative definition of the Balmer jump is given in 
\cite{2005A&A...429..713S}, according to which the Balmer jump is located at the 
wavelength on which the Balmer continuum (lower envelope of spectral lines in 
Fig.~\ref{fig:bj-cross-section}) intersects the Lyman continuum. This alternative 
definition has the advantage to better fit what visually looks like the Balmer jump 
of the intensity spectrum of the Sun, and is located at wavelengths
much longer than our definition (in the cited paper, for solar conditions and taking 
also pressure broadening mechanisms into account, the position is reported to be 
shifted by $140\,\AA$, while a shift of $100\,\AA$ is reported from observational 
data).

We emphasize that the Gaussian broadening displayed in the bottom panel of 
Fig.~\ref{fig:bj-cross-section} is not representative of any specific plasma. 
Its only purpose is to provide a qualitative description of the effect of 
additional broadening. In addition to the thermal Gaussian broadening (Doppler 
broadening), the interaction between the almost-free 
electrons close to the photoionization threshold and the electric field of the 
surrounding ions and electrons in a realistic plasma has to be taken into account 
\citep[see e.g. ][]{1960ApJ...132..883G}. The electric field is not only 
responsible for the usual linear and quadratic Stark broadening (which rapidly 
increases with the principal quantum number of the levels), but it also provides 
an unbounded contribution to the Hamiltonian, which results in the quenching of 
the lines with high upper quantum number $n$. This latter aspect has been studied 
by \cite{1931ZPhy...68..204L}, and it is also described at the end of the chapter 
on the Stark effect in the book of \cite{1957qmot.book.....B}.

\section{Conclusion}

The Balmer jump is produced by a rapid drop of the total Balmer cross-section, 
but this variation is smooth and continuous when both bound-bound and bound-free 
processes are taken into account, and its shape and location are dependent on how 
it is defined and on the broadening mechanisms. A possible definition of the 
position of the Balmer jump is the location at which the distance between 
consecutive spectral lines equals the FWHM of the spectral lines. In this work, 
we have considered an isolated hydrogen atom (no collisions), and we have found 
that this location is shifted about $7\,\AA$ longward of the Balmer limit when 
considering thermal broadening under solar conditions with respect to the ideal 
situation in which only natural broadening comes into play. 

Moreover, we have shown that at specific wavelengths, and in particular at the 
Balmer limit, it is possible to compute the cross-section in a fully analytical 
way. At the Balmer limit, we found this value to be proportional to 
$\alpha a_0^2$, with $\alpha$ being the fine-structure constant, $a_0$ being 
the Bohr radius, and the proportionality constant being the pure mathematical 
constant $\pi^2\, 2^{12}\times 5 / \exp(8)$.

We proposed a method to deal numerically with the accumulation of an infinite 
amount of lines and applied it to the simple case of the hydrogen atom near 
the Balmer limit. We expect this method to be also applicable to the 
accumulation of spectral lines formed by more complex atoms or molecules.
The modelling of the spectrum near the Balmer limit, taking the interactions with 
the surrounding plasma into account, is expected to be significantly more complex 
than the present approach and is therefore left for a future work.

\appendix

\section{Reduced matrix elements of the dipole operator}
\label{sec:reduced_matrix_elements}

The reduced matrix elements $\reducedME{\alpha J}{\vect{d}}{\alpha' J'}$ of the 
dipole operator $\vect{d}=-e_0\vect{r}$ can be expressed in terms of a radial 
integral in a quite general way by assuming that the quantum numbers contained 
in $\alpha$ and $\alpha'$ are eigenvalues of operators commuting with angular 
momentum. We start from the Wigner--Eckart theorem as given in Eq. (2.96) of 
\citet{2004ASSL..307.....L}:
\begin{multline}
\braket{\alpha JM}{T_q^k}{\alpha' J'M'} \\
= (-1)^{J'+k+M}\sqrt{2J+1} \threej{J}{J'}{k}{-M}{M'}{q} \reducedME{\alpha J}{\vect{T}^k}{\alpha' J'},
\label{eqn:wigner-eckart}
\end{multline}
which is inverted using the orthogonality relations of the 3-$j$ symbols:
\begin{equation}
\begin{split}
\reducedME{\alpha J}{\vect{T}^k}{\alpha' J'} &= (-1)^{J'+k+M}\sqrt{2J+1} \\
&\times \sum_{M'q}\threej{J}{J'}{k}{-M}{M'}{q} \braket{\alpha JM}{T_q^k}{\alpha' J'M'},
\end{split}
\end{equation}
and which holds for any $M\in\{-J,-J+1,\dots,J-1,J\}$. By definition, an irreducible 
spherical vector rotates with the same transformation law as the $l=1$ components 
of the spherical harmonics $Y_l^m$. Hence, from the components $(T_x, T_y, T_z)$ 
of an arbitrary Cartesian vector $\vect{T}$ with norm $T$, we can construct a 
spherical vector whose components are given by $c T Y_1^m(\vect{T}/T)$, with $c$ 
an arbitrary constant.

Fixing $c$ provides a one-to-one relation between Cartesian vectors and irreducible 
spherical vectors. For consistency with Eq. \eqref{eqn:bf-einstein-coef}, the 
constant is chosen so that the spherical components of the vector field operator 
$\vect{r}$ are given by Eq. (2.82) of \citet{2004ASSL..307.....L}:
\begin{equation}
r^1_m = \sqrt{\frac{4\pi}{3}}rY_1^m(\hat{\vect{r}}),
\end{equation}
which is inserted in the inverted version of the Wigner--Eckart theorem. After setting $M=0$, the resulting expression reads
\begin{equation}
\begin{split}
\reducedME{\alpha J}{\vect{d}}{\alpha' J'} &= e_0(-1)^{J'+1}\sqrt{2J+1}\   \braket{\alpha}{r}{\alpha'} \\
&\times \sqrt{\frac{4\pi}{3}}\sum_{M'q}\threej{J}{J'}{1}{0}{M'}{q}\ \braket{J0}{Y_1^q(\hat{\vect{r}})}{J'M'}.
\end{split}
\end{equation}
The angular integrals can be computed using Weyl's theorem 
\citep[see e.g. relation (8.8) of][]{2014asrp.book.....L}, resulting in
\begin{equation}
\begin{split}
\reducedME{\alpha J}{\vect{d}}{\alpha' J'} &= e_0(-1)^{J'+1}\sqrt{2J'+1}\ \times  \braket{\alpha}{r}{\alpha'} \\
&\times \threej{J}{J'}{1}{0}{0}{0}\ \sum_{M'q}(2J+1)\threej{J}{J'}{1}{0}{M'}{q}^2.
\end{split}
\end{equation}
Using properties of the 3-$j$ symbols, it is possible to check that the sum in the 
previous relation is equal to $1$, leading to our final result
\begin{equation}
\begin{split}
\reducedME{\alpha J}{\vect{d}}{\alpha' J'} &= e_0(-1)^{J'+1}\sqrt{2J'+1}\,\threej{J}{J'}{1}{0}{0}{0} \\
&\quad\times\ \braket{\alpha}{r}{\alpha'}.
\end{split}
\end{equation}

\section{Asymptotic properties of the Whittaker $M$ function}
\label{sec:whitM_asympt_prop}

We show that
\begin{equation}
\begin{split}
\lim_{\kappa\rightarrow\infty} \kappa^{\mu+\frac{1}{2}}M_{\kappa,\mu}\left(\frac{z}{\kappa}\right) &= \lim_{\kappa\rightarrow\infty} (i\kappa)^{\mu+\frac{1}{2}} M_{i\kappa,\mu}\left(-\frac{iz}{\kappa}\right) \\
&= z^{\mu+\frac{1}{2}}\,_0F_1(1+2\mu;-z)
\end{split}
\label{eqn:whitM-lim}
\end{equation}
for $\mu\in\mathbb{C}$. The hypergeometric function $\,_0F_1$ is defined by the series
\begin{equation}
\,_0F_1(\gamma; z) = \sum_{k=0}^{\infty} \frac{1}{(\gamma)_k}\frac{z^k}{k!}, \quad (\gamma)_k \equiv \gamma(\gamma-1)(\gamma-2)\cdots(\gamma-k+1)
,\end{equation}
and the Whittaker $M$ function is defined by
\begin{equation}
M_{\kappa,\mu}(z) = \exp\left(-\frac{z}{2}\right)z^{\mu+\frac{1}{2}}\,_1F_1\left(\mu-\kappa+\frac{1}{2},1+2\mu;z\right).
\label{eqn:whitM-def}
\end{equation}
The confluent hypergeometric function $\,_1F_1$ is studied in detail in \citet{lebedev1965special} and defined by the hypergeometric series
\begin{equation}
\,_1F_1(\alpha,\gamma;z) = \sum_{k=0}^\infty\frac{(\alpha)_k}{(\gamma)_k}\frac{z^k}{k!}, \quad \alpha,\gamma,z\in\mathbb{C},\quad \gamma\neq 0, -1, -2,\dots
\end{equation}
In particular, $_1F_1$ is an entire function of $z$ and of its parameter $\alpha$, 
and a meromorphic function of $\gamma$ with simple poles on points 
$\gamma=0, -1, -2,\text{etc}.$

We first wish to calculate the asymptotic behaviour of 
$_1F_1\left(a+\kappa,\gamma;\frac{z}{\kappa}\right)$ when 
$\kappa\rightarrow\pm\infty$ for fixed complex numbers $a$ and $z$ and 
$\gamma\neq 0, -1, -2, \text{etc.}$ To this aim, we use the dominated convergence 
theorem applied to the measure space $(\SetN, \Sigma, \mu)$, with 
$\Sigma=\mathcal{P}(\SetN)$ the power set of $\SetN$ and $\mu$ the counting 
measure. With these specific choices, the dominated convergence theorem states 
that this limit can be computed term by term in the series expansion of 
$\,_1F_1$, as long as there exists a bounding function $B(k)$ such that for 
some finite number $K$
\begin{equation}
\left|\frac{(a+\kappa)_k}{(\gamma)_k}\frac{1}{\kappa^k}\right| \leq B(k)\quad \forall\, |\kappa| > K,
\end{equation}
with $B(k)$ independent of $\kappa$, and the series
\begin{equation}
\sum_{k=0}^\infty B(k)\, \frac{|z|^k}{k!} < \infty.
\label{eqn:bounded_series}
\end{equation}
A bounding function like this can be found for $|\kappa|>|a|$, $|a+\kappa|>|\gamma|$ and 
$|a+\kappa|>|\gamma|^2$ (we choose for example $K=1+|a|+|\gamma|+|\gamma|^2$),
\begin{equation}
\begin{split}
\left|\frac{(a+\kappa)_k}{(\gamma)_k}\frac{1}{\kappa^k}\right|
&= \left|\frac{1\left(1+\frac{1}{a+\kappa}\right)\cdots\left(1+\frac{k-1}{a+\kappa}\right)}{1\left(1+\frac{1}{\gamma}\right)\cdots\left(1+\frac{k-1}{\gamma}\right)}\,\frac{\left(1+\frac{a}{\kappa}\right)^k}{\gamma^k}\right| \\
&\leq \sup_{n\in\SetN}\left(\left|\frac{1+\frac{n}{a+\kappa}}{1+\frac{n}{\gamma}}\right|\right)^k\frac{2^k}{|\gamma|^k} \quad (\gamma\neq 0,-1,-2,\dots) \\
&= \sup_{n\in\SetN}\left(\left|\left(\gamma-\frac{\gamma^2}{a+\kappa}\right)\frac{1}{n+\gamma}+\frac{\gamma}{a+\kappa}\right|\right)^k\frac{2^k}{|\gamma|^k} \\
&\leq\sup_{n\in\SetN}\left(
\left(|\gamma|+\frac{|\gamma|^2}{|a+\kappa|}\right)\frac{1}{|n+\gamma|}+\frac{|\gamma|}{|a+\kappa|}\right)^k\frac{2^k}{|\gamma|^k} \\
&\leq\left(\frac{1+|\gamma|}{d(-\gamma,\SetN)}+1\right)^k\frac{2^k}{|\gamma|^k} \\
&=\left(\frac{2}{|\gamma|}\left(1+\frac{1+|\gamma|}{d(-\gamma,\SetN)}\right)\right)^k \\
&\equiv B(k) \equiv cB^k,
\end{split}
\end{equation}
where with this bound $c=1$, and we have defined the distance function for $z\in\mathbb{C}$ and 
$E\subset\mathbb{C}$ to be given by
\begin{equation}
d(z,E) \equiv \inf_{w\in E}(|z-w|).
\end{equation}
With the bound $B(k)$, it is clear that Eq. \eqref{eqn:bounded_series} is satisfied:
\begin{equation}
\sum_{k=0}^\infty B(k)\, \frac{|z|^k}{k!} = c\sum_{k=0}^\infty B^k\, \frac{|z|^k}{k!} = c\cdot\exp(B|z|) < \infty,
\end{equation}
but we note that in the relevant case in which $\gamma\geq 2$ (corresponding to the 
angular momentum $l\geq0$), $K$ could be chosen even greater, and with $c=3/2$, it 
is possible to find a similar bound for which $B$ is arbitrarily close to $1/3$
(this bound is not required for the present proof, but will be useful later). 
The limits, term by term in the series expansion of $_1F_1$, are finally given by
\begin{equation}
\lim_{\kappa\rightarrow\pm\infty}\frac{(a+\kappa)_k}{(\gamma)_k}\frac{1}{\kappa^k} = \frac{1}{(\gamma)_k},
\end{equation}
and by application of the dominated convergence theorem
\begin{equation}
\lim_{\kappa\rightarrow\pm\infty} \,_1F_1\left(a+\kappa,\gamma;\frac{z}{\kappa}\right) = \sum_{k=0}^\infty\frac{1}{(\gamma)_k}\frac{z^k}{k!} \equiv \,_0F_1(\gamma;z).
\label{eqn:1F1_lim_real}
\end{equation}
The same reasoning with the same bound, but substituting $\kappa\rightarrow i\kappa$, 
provides a proof of
\begin{equation}
\lim_{\kappa\rightarrow\pm\infty} \,_1F_1\left(a+i\kappa,\gamma;-\frac{iz}{\kappa}\right) = \,_0F_1(\gamma;z).
\label{eqn:1F1_lim_complex}
\end{equation}
The proof of Eq. \eqref{eqn:whitM-lim} then follows by substituting 
$\kappa\rightarrow -\kappa$ (and $\kappa\rightarrow -i\kappa$, respectively), 
$z\rightarrow -z$, $a\rightarrow\mu+\frac{1}{2}$ and $\gamma\rightarrow 1+2\mu$ in Eqs. 
\eqref{eqn:1F1_lim_real} and \eqref{eqn:1F1_lim_complex}, and inserting the resulting 
limits in Eq. \eqref{eqn:whitM-def}.

We note that choosing $1/3<B<1/2$ further allows us to prove that the wavefunctions of 
the bound/free electrons are bounded independently of $n_u$ and $\varepsilon$ 
and that the wavefunction of the bound electron is (asymptotically) exponentially 
decreasing. It is indeed straight-forward to check with this bound that $_1F_1$ in 
Eq.~\eqref{eqn:whitM-def} grows more slowly than $\exp(z/2)$, allowing an exponential 
decay of the Whittaker $M$ function in Eq.~\eqref{eqn:whitM-def}. Hence the 
integrands inside the radial integrals \eqref{eqn:radial_integral_bound} and 
\eqref{eqn:radial_integral_free} are also bounded and (asymptotically) exponentially 
decreasing. Using again the dominated convergence theorem, we can further justify the 
interchange of the limits $n_u\rightarrow\infty$ (and $\varepsilon\rightarrow 0$ 
respectively) and the radial integrals.

\begin{acknowledgements}
We are grateful to Edgar S. Carlin for his help with the computation of the 
reduced matrix elements of the dipole operator, and to Jhih-Huang Li for his 
revision of the derivations of the asymptotic properties of the Whittaker $M$ 
function. We are also very grateful to Ji\v{r}\'i Št\v{e}p\'an and Roberto Casini 
for a preliminary reviewing of this work and for providing new ideas and useful 
feedback, as well as to Javier Trujillo Bueno for inspiring discussions. We also 
acknowledge the referee's valuable and encouraging input.
This work was supported by the Swiss National Science Foundation under grant 
ID~200020\_157103/1.
\end{acknowledgements}

\noteadd We call the reader’s attention to the fact that \cite{1935MNRAS..96...77M} and references therein studied the limit $n_u \rightarrow \infty$ for the wavefunction of the bound electron and came to the conclusion that there was exact continuity of the absorption coefficient at the Balmer limit. In the present paper, we used the modern formalism of \cite{2004ASSL..307.....L} to prove continuity.

\bibliographystyle{aa}
\bibliography{aa31974-17}
\balance

\end{document}